# Electronic structure of nanorod of strongly prolate ellipsoidal shape

T. Kereselidze, T. Tchelidze and T. Nadareishvili

Faculty of Exact and Natural Sciences, Tbilisi State University, Chavchavadze Avenue 3, 0179 Tbilisi, Georgia

**Abstract.** A charged particle (electron or hole) confined in nanorod of strongly prolate ellipsoidal shape is considered. The effective-mass Schrödinger equation is solved in prolate spheroidal coordinates and asymptotically exact expressions for the energy spectrum and wavefunctions are derived. It is shown that the treatment of the confinement energies is incorrect if the actual shape of ellipsoidal nanorod is not taken into account. In particular, developed earlier the approach based on the consideration of the problem in cylindrical coordinates with the use of parabolic potential leads to the incorrect expression for the energy spectrum.

Rule of correlation between the states corresponding to spherical quantum dot and nanorod of strongly prolate ellipsoidal shape is suggested and an appropriate energy correlation diagram is constructed. The correlation diagram is in complete qualitative and quantitative agreement with the energy diagram obtained by numerical solution of the Schrödinger equation in spheroidal coordinates.

## 1. Introduction

In the last decade there is a rapidly growing interest in nanosized crystalline semiconductor structures of various shapes. Physical properties of nanosized objects such as nanorods, nanowires and quantum dots have been intensively investigated both theoretically and experimentally [1]. The most important result that has been revealed in these investigations is the strong interdependence between the character of energy spectrum of nanosized object and its geometrical parameters, such as size and shape. Size and shape tunable control of spectral and optical characteristics of nanoobjects opens exciting possibilities for the creation of new functional materials that can be used in unlimited application.

From the theoretical point of view, spherical quantum dots are the easier to investigate. Because of the symmetry spherical quantum dots allow us to obtain analytical solution for the energy spectrum, coefficient of absorption, charge carriers mobility etc. [2,3]. On the other hand, modern growth techniques make possible to obtain nanoobjects of different geometrical shapes and sizes. In the literature there are many works devoted to the theoretical study of cylindrical, pyramidal, ellipsoidal, semiellipsoidal nanosized objects [4-18].

For nanorod of cylindrical shape the problem can be solved exactly assuming that the potential, which is experienced by a charged particle (electron or hole), is zero inside the cylinder and that it is infinite outside. The energy for the particle confined to a cylinder is represented as a sum of two terms, where the first term reflects the radial contribution to the total energy (the motion perpendicular to the cylinder axis) and the second one represents the small longitudinal energy contribution [4].

Obviously, the first attempt to consider a particle trapped inside an ellipsoid has been done in [5]. In this work has been assumed that a spherical potential well with infinitely high walls is subject to a small deformation, which gives it the form of a slightly prolate or oblate ellipsoid. Accordingly, the splitting of the energy levels of a particle has been found using perturbation theory. In [6] this approach has been applied for finding the Fermi energy of ellipsoidal nanoparticles.

For nanorod that has the shape of an ellipsoid of revolution with major semi-axis $c$ (the nanorod axis) much larger than minor semi-axes $a = b$ (the nanorod radius), the problem is usually treated in cylindrical coordinates by using an adiabatic approximation [7-12]. The significant elongation of nanorod ($c \gg a$) makes it possible, by employing the adiabatic approximation, to separate the parallel motion from the motion perpendicular to the nanorod axis. Within this approximation, the "slow" motion of an electron (or hole) parallel to the nanorod axis is initially neglected and the spectrum of the electron in two dimension confinement perpendicular to the nanorod axis is found. The parallel motion is treated by averaging the Hamiltonian over the "fast" motion of the electron strongly confined in two dimensions.

In nanorods the electron states are classified by the absolute values of the angular momentum projection on the nanorod axis $m = 0, 1, 2, \ldots$. The wavefunction of the electron motion perpendicular to the nanorod axis can be represented as $J_m(k\rho)e^{\pm im\varphi}$, where $\varphi$ is the azimuthal angle, $\rho$ is the distance from the nanorod axis, and $J_m(k\rho)$ is Bessel function of the $m$th order. The momentum $\hbar k$ is related to the kinetic energy of the electron motion perpendicular to the nanorod axis, $E = \hbar^2 k^2 / 2m^*$ where $m^*$ is the electron effective mass. From the condition that the wavefunction vanishes at the nanorod surface, one can find the momentum $\hbar k$ which satisfies the boundary condition, $\hbar k_{s,m} = \hbar \tau_{s,m}/a$ where $\tau_{s,m}$ is the sth root of Bessel function $J_m(k\rho)$. This gives $E_{s,m} = \hbar^2 \tau_{s,m}^2/(2m^* a^2)$ for the spectrum connected to the "fast" motion of electron perpendicular to the nanorod axis [7]. The energy levels connected to the motion of electron parallel to the nanorod axis are obtained by employing the Hamiltonian that is averaged over the "fast" motion. Thus each level of the "fast" subsystem has its own family of levels, which are created by the "slow" subsystem.

In [8-12] the energy expression for the "fast" subsystem has been expanded in Taylor series and the obtained parabolic potential has been used for the treating the "slow" subsystem. The developed calculating scheme leads to a set of equidistant energy levels for the motion of electron along the nanorod axis. In [8-12] it is stated that the obtained result is valid only for the low spectrum levels, i.e. for the small quantum numbers. Substitution of parabolic potential by Pöschl-Teller potential [19] does not remove the weakness of developed approach, because Pöschl-Teller potential practically coincides with parabolic potential for the low-lying levels. The correct solution of the problem shows that the expression derived in [8-12] for the energy spectrum of the "slow" subsystem is incorrect, and accordingly the energy levels are not equidistant.

The difficulties arisen in [8-12] are connected with the consideration of nanorod that has the shape of an ellipsoid in cylindrical coordinates. Obviously, spheroidal coordinates [20] contain the symmetry that is intrinsic to nanorod of ellipsoidal shape. Therefore, spheroidal coordinates are the most appropriate for the analytical solution of the problem. Analyzing the results obtained in [13-18] by means of numerical solving of the problem in spheroidal coordinates, this statement becomes evident. In the regime of strong size quantization, the electron-hole Coulomb interaction energy is much less than the confinement energy. Therefore, in this regime one can neglect the Coulomb interaction and to treat the electron and the hole independently.

In the present work we consider nanorod of strongly prolate ellipsoidal form. Using spheroidal coordinates in which the Schrödinger equation is separable, we show that the obtained differential equations can be solved analytically and the asymptotically exact expressions for the energy spectrum and wavefunctions can be obtained. A small parameter that is used for finding the asymptotic solutions is an inverse value of a distance between the foci of prolate spheroidal coordinates. Thus the purpose of the present study is to show that the calculation of the confinement energies is incorrect if the actual shape of ellipsoidal nanorod is not taken into account.

The paper is organized as follows. After a brief review of the main results of previous authors and stating the goal of the present study (section 1), the basic equations are given in prolate spheroidal coordinates (section 2). In sections 3, we derive the analytical expressions for the energy and wavefunctions of an electron confined in ellipsoid. In section 4 a rule of correlation between the states corresponding to spherical quantum dot and nanorod of strongly prolate ellipsoidal shape is suggested, and an appropriate energy correlation diagram is constructed and compared with an energy diagram obtained by numerical solution of the Schrödinger equation in spheroidal coordinates.

## 2. Basic equations

Let us solve the model problem of an electron in a strongly prolate ellipsoid. The surface of the ellipsoid is defined by equation

$$\frac{x^2}{a^2} + \frac{y^2}{b^2} + \frac{z^2}{c^2} = 1 \tag{1}$$

where $a = b$ are the small and $c$ is the big semi-axes. In the ellipsoid, we have the geometry shown in figure 1 and the associated boundary conditions. Namely, the potential inside the ellipsoid is zero and that it is infinity outside. As a consequence, the electron is confined inside the finite-volume ellipsoid.

Evidently, prolate spheroidal coordinates $\xi, \eta, \varphi$ ($1 \leq \xi < \infty, -1 \leq \eta \leq 1, 0 \leq \varphi < 2\pi$)

$$\begin{aligned} x &= \frac{R}{2}\sqrt{(\xi^2 - 1)(1 - \eta^2)} \cos\varphi \\ y &= \frac{R}{2}\sqrt{(\xi^2 - 1)(1 - \eta^2)} \sin\varphi \\ z &= \frac{R}{2}\xi\eta \end{aligned} \tag{2}$$

with the foci $z = \pm R/2$, where $R$ is the confocal distance along the $z$ - axis, are the most appropriate coordinates for the solution of the problem.

In speroidal coordinates (1) is converted into the equation

$$\frac{(\xi^2 - 1)(1 - \eta^2)}{a^2} + \frac{\xi^2 \eta^2}{c^2} = \frac{4}{R^2}. \tag{3}$$

By putting $\eta = \pm 1$ in (3), it can be easily found that $\xi$ becomes

$$\xi^* = \frac{2c}{R} = \frac{1}{e} \tag{4}$$

at the surface of ellipsoid where the potential is infinity. In (4) $e$ is the eccentricity of the ellipse generating the constant $\xi^*$ surface, which is ellipsoid.

Our purpose is to solve the effective-mass Schrödinger equation

$$\left(\Delta + k^2\right)\Psi = 0 \tag{5}$$

in spheroidal coordinates and to find the electron energies and the corresponding wavefunctions. In equation (5) $k^2 = 2m^* E/\hbar^2$, where $m^*$ is the effective mass of the electron and $E$ is the electron energy which is measured from the bottom of the well.

Representing the full wavefunction $\Psi$ as a product of three functions

$$\Psi = X(\xi)Y(\eta)\frac{e^{\pm im\varphi}}{\sqrt{2\pi}} \tag{6}$$

we arrive at the two coupled equations after separation of variables in (5)

$$\frac{d}{d\xi}(\xi^2 - 1)\frac{dX}{d\xi} + \left[\lambda + \frac{k^2 R^2}{4}(\xi^2 - 1) - \frac{m^2}{\xi^2 - 1}\right]X = 0 \tag{7.a}$$

$$\frac{d}{d\eta}(1 - \eta^2)\frac{dY}{d\eta} + \left[-\lambda + \frac{k^2 R^2}{4}(1 - \eta^2) - \frac{m^2}{1 - \eta^2}\right]Y = 0. \tag{7.b}$$

Here $\lambda$ is the separation constant, $X(\xi)$ and $Y(\eta)$ denote the "quasiradial" and "quasiangular" wavefunctions by analogy to the spherical quantum dot [1].

Instead of functions $X(\xi)$ and $Y(\eta)$ new functions [20]

$$\begin{aligned} U(\xi) &= (\xi^2 - 1)^{1/2} X(\xi) \\ V(\eta) &= (1 - \eta^2)^{1/2} Y(\eta) \end{aligned} \tag{8}$$

are usually introduced to find the asymptotic solutions of equations (7). These new functions have to satisfy the boundary conditions

$$U(1) = 0 \quad U(\xi^*) = 0 \tag{9.a}$$

$$V(\pm 1) = 0. \tag{9.b}$$

Here $U(\xi^*) = 0$ reflects the condition that the electron is trapped inside the ellipsoid, equations $U(1) = 0$ and $V(\pm 1) = 0$ provide the condition that the electron wavefunctions $X(\xi)$ and $Y(\eta)$ must be finite quantity at $\xi = 1$ and $\eta = \pm 1$.

Substituting (8) into (7) we come to the following equations for the unknown functions $U(\xi)$ and $V(\eta)$

$$\frac{d^2 U}{d\xi^2} + \left[ \frac{k^2 R^2}{4} + \frac{\lambda}{\xi^2 - 1} + \frac{1 - m^2}{(\xi^2 - 1)^2} \right] U = 0 \qquad (10.\text{a})$$

$$\frac{d^2 V}{d\eta^2} + \left[ \frac{k^2 R^2}{4} - \frac{\lambda}{1 - \eta^2} + \frac{1 - m^2}{(1 - \eta^2)^2} \right] V = 0. \qquad (10.\text{b})$$

It follows from (10.a) and (10.b) that these equations have irregularities of the first order and the second order at $\xi = 1$ and $\eta = \pm 1$. Another peculiarity is that (10.a) and (10.b) (as well as (7.a) and (7.b)) are identical equations defined in the different regions.

## 3. Asymptotically exact solutions

We assume that the small semi-axes of strongly prolate ellipsoid are equal to few nanometers whereas the big semi-axis $c \gg a$. In this case the confocal distance $R$, that appears in equations (7) and (10), is a big value $R = 2\sqrt{c^2 - a^2} \gg a$. Accordingly, the inverse value $R^{-1}$ is small and therefore, this parameter can be used for finding the asymptotic solutions of equations (10.a) and (10.b).

At first we consider more complicated "quasiangular" equation (10.b) and determine separation constant $\lambda$. We assume that for $\eta \to 0$ the eigenfunctions of equation (10.b) convert into the wavefunctions describing the motion parallel to the $z$-axis in nanorod of cylindrical shape. This assumption is evident, because away from the focuses namely, in the region $\eta \sim 0$ the surface of strongly prolate ellipsoid slightly differs from the surface of cylinder. In appendix A it is shown that along the $z$-axis the motion of a charged particle confined in a cylinder with a large but finite length $L = R\xi^*$ is described by the two sets of wavefunctions (A.2) and (A.3).

Let us introduce the new variable $z = R\xi^*\eta/2$ and rewrite equation (10.b) in the following form

$$\frac{d^2 V}{dz^2} + \left[ p^2 + \upsilon(z) \right] V = 0 \qquad (11.\text{a})$$

were

$$p = \frac{k}{\xi^*} \left( 1 - \frac{4(\lambda + m^2 - 1)}{k^2 R^2} \right)^{1/2}$$

and  (11.b)

$$\upsilon = \frac{16 z^2}{R^2 \xi^{*2}} \left( \frac{\lambda}{4z^2 - R^2 \xi^{*2}} - \frac{2(1 - m^2)(2z^2 - R^2 \xi^{*2})}{(4z^2 - R^2 \xi^{*2})^2} \right).$$

The second term in the square brackets in (11.a) is a small value. Exceptions are the regions near the left- and right-hand focuses where $z \simeq \pm R\xi^*/2$. Excluding these regions from the consideration and neglecting the term $\upsilon(z)$, the general solution of equation (11.a) can be represented as follows

$$V(z) = c_1 e^{ipz} + c_2 e^{-ipz} \qquad (12)$$

where $c_1$ and $c_2$ are the constant coefficients.

It is clear that if $c_1 = c_2 = A/2$ and $p = (2n+1)\pi / R\xi^*$, (12) converts into (A.2) and if $c_1 = -c_2 = -iB/2$ and $p = 2n\pi / R\xi^*$, (12) converts into (A.3). These conditions allow us to determine the separation constant in equation (10.b). Small calculation gives

$$\lambda^{(+)} = \frac{1}{4}\left(k^2 R^2 - \pi^2(2n+1)^2 + 4(1-m^2)\right) \qquad (13.a)$$

for the symmetrical and

$$\lambda^{(-)} = \frac{1}{4}\left(k^2 R^2 - 4\pi^2 n^2 + 4(1-m^2)\right) \qquad (13.b)$$

for the antisymmetrical states with respect to reflection in the plane normal to and bisecting the nanorod axis (replacement $\eta \to -\eta$ or alternatively $z \to -z$). In (13.a) $n = 0, 1, 2, \ldots$ and in (13.b) $n = 1, 2, 3, \ldots$, as it is indicated in appendix A.

Now we solve the "quasiradial" equation in (10) using the obtained values for the separation constant. Introducing the new variable $\rho = ikR(\xi - 1)$ ($0 \leq \rho < \infty$), equation (10.a) can be rewritten in the form

$$\frac{d^2 U^{(\pm)}}{d\rho^2} + \left[-\frac{1}{4} + \frac{\lambda^{(\pm)}}{2ikR\rho(1+\rho/2ikR)} + \frac{1-m^2}{4\rho^2(1+\rho/2ikR)^2}\right] U^{(\pm)} = 0 \quad (14)$$

where $\lambda^{(\pm)}$ are defined by expressions (13.a) and (13.b).

Making the identical transformations [21]

$$\frac{1}{1+\rho/2ikR} = 1 - \frac{\rho/2ikR}{1+\rho/2ikR}$$

$$\frac{1}{(1+\rho/2ikR)^2} = 1 - \frac{\rho/ikR + (\rho/2ikR)^2}{(1+\rho/2ikR)^2} \qquad (15)$$

we arrive at the equation

$$\frac{d^2 U^{(\pm)}}{d\rho^2} + \left[-\frac{\alpha^{(\pm)2}}{4} + \frac{\beta^{(\pm)}}{ikR\rho} + \frac{1-m^2}{4\rho^2} + \omega^{(\pm)}(\rho)\right] U^{(\pm)} = 0 \qquad (16.a)$$

where

$$\alpha^{(\pm)} = \left(1 - \frac{8\beta^{(\pm)} + m^2 - 1}{4k^2 R^2}\right)^{1/2}$$

$$\omega^{(\pm)} = \frac{\rho}{4(ikR)^3}\left(1+\frac{\rho}{2ikR}\right)^{-1}\left[\beta^{(\pm)} - \frac{1-m^2}{4}\left(1+\frac{\rho}{4ikR}\right)\left(1+\frac{\rho}{2ikR}\right)^{-1}\right] \qquad (16.b)$$

and $\beta^{(\pm)} = (2\lambda^{(\pm)} + m^2 - 1)/4$.

In equation (16.a) the last term in the square brackets is proportional to $R^{-1}$ inside the ellipsoid. Therefore, this term can be neglected in (16.a) as a small one. The solution of the residual equation, which satisfies the necessary boundary condition at $\rho = 0$, is the Whittaker function [22]. Thus returning to the variable $\xi$, we obtain for the solution of equation (10.a)

$$U^{(\pm)} = e^{-\frac{ikR}{2}(\xi-1)\alpha^{(\pm)}}(\xi-1)^{\frac{m+1}{2}} F\left(-\frac{\beta^{(\pm)}}{ikR\alpha^{(\pm)}} + \frac{m+1}{2}, m+1, ikR(\xi-1)\alpha^{(\pm)}\right) + O(R^{-1}) \quad (17)$$

where $F(\alpha, \beta, x)$ is the confluent hypergeometric function.

The variable of the confluent hypergeometric function in (17) is restricted inside the ellipsoid, whereas the first parameter is large $\left(\beta^{(\pm)} R^{-1} \sim R \gg 1\right)$. This fact allows us to use the asymptotic expansion for the confluent hypergeometric function with respect to the first parameter [23]. In consequence, we obtain for the "quasiradial" wavefunctions

$$U^{(\pm)} = m!\left(\beta^{(\pm)}\right)^{-m/2}(\xi-1)^{1/2} J_m\left(2\sqrt{\beta^{(\pm)}(\xi-1)}\right) + O(R^{-1}) \qquad (18)$$

where $J_m(x)$ is Bessel function of the first kind.

It follows from the boundary condition (9.a) that wavefunctions (18) must be equal to zero on the surface of the ellipsoid. Hence, we have

$$\beta^{(\pm)}(\xi^* - 1) = \frac{\tau_{s,m}^2}{4}. \tag{19}$$

Here $\xi^*$ is defined by expressions (4) and $\tau_{s,m}$ are the dual-index Bessel function roots $J_m(\tau_{s,m}) = 0$. Index $s$ enumerates the roots of Bessel function.

Condition (19) through equation $E = \hbar^2 k^2 / 2m^*$ gives for the electron energy

$$E_{s,n,m}^{(\pm)} = \frac{\hbar^2}{2m^*} \left[ \frac{\tau_{s,m}^2}{2\left(c - \sqrt{c^2 - a^2}\right)\sqrt{c^2 - a^2}} + \frac{\pi^2 n^{(\pm)2}}{4(c^2 - a^2)} + \frac{m^2 - 1}{2(c^2 - a^2)} \right] \tag{20}$$

where $n^{(+)} \equiv 2n+1 = 1, 3, \ldots$ for the symmetrical and $n^{(-)} \equiv 2n = 2, 4 \ldots$ for the antisymmetrical states.

Now we turn to equation (10.b) and determine the "quasiangular" wavefunctions in the whole region of definition of $\eta$. Expression in the square brackets may be expanded into elementary fractions and equation (10.b) can be represented in the following form [24]

$$\frac{d^2 V}{d\eta^2} + \left[ \frac{k^2 R^2}{4} - \frac{\beta}{(1+\eta)} + \frac{1-m^2}{4(1+\eta)^2} - \frac{\beta}{(1-\eta)} + \frac{1-m^2}{4(1-\eta)^2} \right] V = 0 \tag{21}$$

where $\beta = (2\lambda + m^2 - 1)/4$.

Introducing the new variable $\rho_1 = ikR(1+\eta)$ ($0 \le \rho_1 \le 2ikR$) we arrive at the equation

$$\frac{d^2 V}{d\rho_1^2} + \left[ -\frac{1}{4} - \frac{\beta}{ikR\rho_1} + \frac{1-m^2}{4\rho_1^2} + \frac{\beta}{2k^2 R^2 (1 - \rho_1/2ikR)} - \frac{1-m^2}{16k^2 R^2 (1 - \rho_1/2ikR)^2} \right] V = 0. \tag{22}$$

Excluding the region near the right-hand focus from the consideration ($\rho_1 \simeq 2ikR$) and neglecting the last two terms in the square brackets in (22), we get

$$\frac{d^2 V^{(0)}}{d\rho_1^2} + \left[ -\frac{1}{4} - \frac{\beta}{ikR\rho_1} + \frac{1-m^2}{4\rho_1^2} \right] V^{(0)} = 0. \tag{23}$$

Solution of equation (23) regular at $\rho_1 = 0$ is the Whittaker function [22]

$$V^{(0)} = e^{-\frac{ikR}{2}(1+\eta)} \left(ikR(1+\eta)\right)^{\frac{m+1}{2}} F\left(\frac{\beta}{ikR} + \frac{m+1}{2}, m+1, ikR(1+\eta)\right) \tag{24}$$

where $F(\alpha, \beta, x)$ is the confluent hypergeometric function.

At $\eta \to 1$ the behavior of $V(\eta)$ is determined by the last term in the square brackets in (21). The solution, which satisfies the necessary condition of finiteness at $\eta = 1$ (see equation (9.b)) is proportional to $(1-\eta)^{(m+1)/2}$. Therefore, it is natural to seek the solution of equation (21) in the form

$$V(\eta) = V^{(0)}(\eta)(1-\eta)^{\frac{m+1}{2}} f(\eta) \tag{25}$$

where $V^{(0)}(\eta)$ is defined by expression (24) and $f(\eta)$ is the unknown function to be determined.

Substituting (25) into (21) and taking into consideration (24), one can find that $f(\eta)$ satisfies the equation

$$\frac{d^2 f}{d\eta^2} - \left[ ikR - \frac{m+1}{1+\eta} - \frac{2}{F} \frac{dF}{d\eta} + \frac{m+1}{1-\eta} \right] \frac{df}{d\eta}$$
$$- \frac{1}{1-\eta} \left[ \beta - \frac{m+1}{2} \left( ikR - \frac{m+1}{1+\eta} - \frac{2}{F} \frac{dF}{d\eta} \right) \right] f = 0. \tag{26}$$

The obtained equation is exact, because at its determination no approximation has been made. Obviously, the solution of equation (26) should satisfy the boundary condition: $f(\eta)$ tends to unity for $\eta \to -1$.

We are interested in the solution of equation (26) in the region near the right-hand focus ($\eta \simeq 1$). In this region derivative $dF/d\eta$ can be calculated using asymptotic expression for $F$ (see appendix B). Introducing the new variable $\rho_2 = ikR(1-\eta)$ and neglecting the terms less than $O(R^0)$ in (26), we come to the confluent hypergeometric equation

$$\rho_2 \frac{d^2 f}{d\rho_2^2} + (m+1-\gamma\rho_2)\frac{df}{d\rho_2} - \left(\frac{\beta}{ikR} + \frac{m+1}{2}\gamma\right)f = 0 \qquad (27.a)$$

where

$$\gamma = 1 - \frac{\beta}{k^2 R^2} + \left(\frac{2}{G}\frac{dG}{d\eta}\right)_{\eta=1}. \qquad (27.b)$$

The solution of equation (27.a), which tends to unity for $\eta \to -1$ is

$$f = \frac{F\left(\frac{\beta}{ik\gamma R} + \frac{m+1}{2}, m+1, ik\gamma R(1-\eta)\right)}{F\left(\frac{\beta}{ik\gamma R} + \frac{m+1}{2}, m+1, 2ik\gamma R\right)} + O(R^{-1}) \qquad (28)$$

Substituting (24) and (28) into (25), we obtain the "quasiangular" wavefunctions which are valid in the whole region of definition of $\eta$. We note that the derived wavefunctions satisfy the necessary boundary conditions at $\eta = \pm 1$.

Replacement $\eta \to -\eta$ leaves equation (21) as well as equations (10.b) and (7.b) unaltered. This means that the wavefunctions $V(-\eta)$ are solutions of equation (21) together with the wavefunctions $V(\eta)$. The full unnormalized "quasiangular" wavefunctions are thus the sum and difference of wavefunctions $V(\eta)$ and $V(-\eta)$ [7]

$$V^{(\pm)}(\eta) = V(\eta) \pm V(-\eta) \qquad (29.a)$$

where

$$V(\pm\eta) = e^{-\frac{ikR}{2}(1\pm\eta)}(1-\eta^2)^{\frac{m+1}{2}} F\left(\frac{\beta}{ikR} + \frac{m+1}{2}, m+1, ikR(1\pm\eta)\right)$$
$$\cdot F\left(\frac{\beta}{ik\gamma R} + \frac{m+1}{2}, m+1, ik\gamma R(1\mp\eta)\right) \qquad (29.b)$$

and $\beta = \beta^{(\pm)}$. It is clear that $V^{(+)}(\eta)$ are symmetrical and $V^{(-)}(\eta)$ are antisymmetrical wavefunctions with respect to the replacement $\eta \to -\eta$. For $\eta \sim 0$ $V^{(+)}(\eta)$ converts into the sum and $V^{(-)}(\eta)$ converts into the difference of plane waves.

**4. Results and discussion**

In section 3 we have obtained the wavefunctions and energies for the electron confined in a prolate ellipsoid. The derived expressions for the energy spectrum and wavefunctions are valid for an ellipsoid with the major semi-axis $c$ much larger than the minor semi-axes $a = b$. The expressions obtained for the energy spectrum as well as for the wavefunctions are asymptotically exact. The main assumption, which has been employed at the derivation, is that in the region $\eta \sim 0$ the wavefunctions corresponding to the motion parallel to the nanorod axis covert into the wavefunctions describing the motion along the $z$- axis in nanorod of cylindrical shape.

It is seen from expression (20) that the energy for the electron confined in nanorod of strongly prolate ellipsoidal shape is a sum of three terms, where the first term reflects the "radial" contribution to the total energy and the second and third ones represent the small energy contribution corresponding to the motion parallel to the nanorod axis. Another peculiarity is that the electron energies in units of ($\hbar^2/2m^*a^2$), depend on the spheroidal aspect ratio $c/a$ but not on $a$ or $c$ separately. Calculation shows that if (20) is expanded in terms of small value $c/a$ and the terms proportional to $(c/a)^2$ and less are neglected, the first and second terms in (20) convert into expression (A.4) that is derived for the energy spectrum of nanorod of cylindrical shape.

It follows from expression (20) that the second term in the squared brackets is proportional to $(2n+1)^2$ ($n=0,1,2,...$) for the symmetrical states and it is proportional to $(2n)^2$ ($n=1,2,...$) for the antisymmetrical states. Hence the energy levels connected to the motion of electron along the nanorod axis are not equidistant. It means that the assumption, that the electron is mainly localized in the region $|z| \ll c$ ($\eta \sim 0$) made in [8-12] and which leads to the equidistant energy levels is not correct. Estimation of the probability to find electron far away from the origin of coordinate system, i.e. in the regions $\eta \sim \pm 1$ by means of derived "quasiangular" wavefunctions (29.a) and (29.b), makes this statement evident.

For an electron confined in slightly prolate ellipsoid with semi-axes $a=b$ and $c \simeq a$, the energy spectrum is defined by the expression [5]

$$E_{N,l,m} = \frac{\hbar^2}{2m^*} \frac{\tau_{N,l}^2}{a^2} \left[ 1 - \frac{c-a}{c+2a} \Delta_{l,m} \right] \quad (30.a)$$

where

$$\Delta_{l,m} = \frac{4}{(2l-1)(2l+3)} \left( l(l+1) - 3m^2 \right). \quad (30.b)$$

In (30.a) $l$ is the angular momentum of the electron, $N=1,2,3,...$ numbers the energy levels in the spherical well for a given $l$, which are independent of $m$ and $\tau_{N,l}$ are the roots of spherical Bessel function $j_l(\tau_{N,l}) = 0$.

So for nanorod of strongly prolate ellipsoidal shape the energy spectrum is defined by expressions (20), whereas for the electron confined in a slightly prolate ellipsoid the energy spectrum is given by expressions (30). In figure 2 the behavior of the energy terms corresponding to the ground and seven excited states, when $a$ is fixed and $c$ increases from $c=a$ (spherical quantum dot) to $c=8a$ (nanorod of strongly prolate ellipsoidal shape), is shown. The energy terms are calculated using expression (20) for $c > 3a$ and expression (30) for $c \simeq a$.

In the spherical quantum dot limit the electronic states are specified by the quantum numbers ($N,l,m$). In the opposite limit, i.e. in nanorod of strongly prolate ellipsoidal shape, the electronic states are classified by the quantum numbers [$s,n,m$]. We assume that correlation between these two sets of states occurs according to conservation of symmetry. Namely, we suppose that there is a one-to-one correspondence between the states with same $m$. Moreover, since the symmetry of the angular wavefunctions is given by $(-1)^{l-m}$ in spherical quantum dot [5], when $c$ increases from $c=a$ to $c \gg a$ the states with even value of $l-m$ transform into the symmetrical states, whereas the sates with odd $l-m$ convert into the antisymmetrical states of nanorod of ellipsoidal shape. The same assumption can be made when $c$ is fixed and $a$ tends to $c$.

Thus the following correlation rule can be formulated: starting from the state with a given symmetry in spherical quantum dot and increasing ellipsoid aspect ratio $c/a$, we come to the lowest not yet occupied (correlated) state of nanorod of strongly prolate ellipsoidal shape, which has the same symmetry. This rule is in complete agreement with the behavior of energy terms obtained in [15] by the numerical solution of the Schrödinger equation in spheroidal coordinates. It should be noted here that the suggested correlation rule is analogous to the molecular orbital correlation rule in the $H_2^+$-like molecular ions [25,26]. The energy terms presented in figure 2 are plotted according to the above formulated correlation rule (see the dashed curves in figure). For $c > 4a$, the energy terms calculated using expression (20) are in good quantitative agreement with the results obtained in [15].

Figure 2 shows that the spherical quantum dot degeneracy with respect to $m$ is removed, as effect of the loss of spherical symmetry. At the same time, the states with the same $N$ and $m$ but different $l$ become almost degenerate at large values of $c/a$. An explanation can be given by considering that if at fixed $a$ we increase $c$, each ellipsoid confined state changes continuously in such a way that its limit is just a cylindrical quantum wire state having the same $m$ and the same number of nodes of the wavefunction along the "radial" coordinate. All the ellipsoid quantum states with the same $N$ and $m$ but different $l$ create a miniband of the cylindrical quantum wire, which explains why, for instance, the states with $N=1$, $m=0$ and $l=0,1,2,\ldots$ become nearly degenerate for $c/a \gg 1$.

Another peculiarity that follows from figure 2 is appearance of levels crossings. Analysis shows that a number of levels crossings increases if the excited states are involved into consideration. An explanation is that, the energy levels with the same symmetry are differently ordered in spherical quantum dot and in nanorod of strongly prolate ellipsoidal shape that leads to the existence of levels crossings. Appearance of levels crossings is interesting because, together with the zeros of dipole moments, it can produce a structure [27] in the profiles of spectral lines emitted from ensemble of nanorods.

In conclusion, we would like to emphasize that comparison with the results obtained by the numerical solution of the problem shows that, the expression derived in the present study for the energy spectrum is valid for different nanosized objects having ellipsoidal shape with the spheroidal aspect ratio $c/a > 4$.

**Acknowledgment**

This work has been supported by Science and Technology Center of Ukraine (Grant N 5633) and Rustavely National Science Foundation (Grant N 09/15) (Georgia).

**Appendix A**

Let us consider an electron confined in a cylinder with a large but finite length $L$. After separation of variables in cylindrical coordinates, the solution of equation describing the motion of the electron along the $z$-axis can be represented as a linear combination of plane waves [4]

$$\Phi(z) = a e^{i\sigma z} + b e^{-i\sigma z}. \tag{A.1}$$

Here $a$ and $b$ are the constant coefficients and $\sigma$ is the separation constant which should be determined.

By using the boundary conditions $\Phi(L/2) = \Phi(-L/2) = 0$, we obtain that in (A.1) $a = b$ and $a = -b$. The corresponding separation constants are, for $a = b$ $\sigma = (2n+1)\pi/L$ with $n = 0,1,2,\ldots$ and for $a = -b$ $\sigma = 2n\pi/L$ with $n = 1,2,3,\ldots$.

We thus obtain the two sets of solutions: symmetrical

$$\Phi^{(+)}(z) = A \cos\left(\frac{(2n+1)\pi}{L} z\right) \tag{A.2}$$

and antisymmetrical

$$\Phi^{(-)}(z) = B \sin\left(\frac{2n\pi}{L} z\right) \tag{A.3}$$

with respect to the replacement $z \to -z$. For the appropriate electronic energies we can write *

$$E^{(\pm)}_{s,n,m} = \frac{\hbar^2}{2m^*}\left[\frac{\tau^2_{s,m}}{a^2} + \frac{\pi^2 n^{(\pm)2}}{L^2}\right] \tag{A.4}$$

where $n^{(+)} = 2n+1$ and $n^{(-)} = 2n$.

**Appendix B**

For large $\rho_1$ the confluent hypergeometric function $F(\beta/ikR+(m+1)/2, m+1, \rho_1)$ has the asymptotic form [23]

---
*In [4] the origin of coordinate system is not located in the center of cylinder; therefore, the electronic states are not divided into symmetrical and antisymmetrical ones (see equation (7.27)).

$$F\left(\frac{\beta}{ikR}+\frac{m+1}{2},m+1,\rho_1\right) = m!\left[\frac{e^{\rho_1}\rho_1^{\frac{\beta}{ikR}-\frac{m+1}{2}}}{\Gamma\left(\frac{m+1}{2}+\frac{\beta}{ikR}\right)}G\left(\frac{m+1}{2}-\frac{\beta}{ikR},-\frac{m-1}{2}-\frac{\beta}{ikR},\rho_1\right)\right.$$
$$\left.+\frac{(-\rho_1)^{-\frac{\beta}{ikR}-\frac{m+1}{2}}}{\Gamma\left(\frac{m+1}{2}-\frac{\beta}{ikR}\right)}G\left(\frac{m+1}{2}+\frac{\beta}{ikR},-\frac{m-1}{2}+\frac{\beta}{ikR},-\rho_1\right)\right]$$

(B.1)

where

$$G(a,b,x) = 1 + \frac{ab}{1!x} + \frac{a(a+1)b(b+1)}{2!x^2} + \cdots$$

(B.2)

and $\Gamma(x)$ is the gamma function.

**References**

1. Harrison P 2005 *Quantum Well, Wires and Dots* (Wiley, New York)
2. Efros Al L and Efros A L 1982 *Sov. Phys. Semicond.* **16** 772
3. Ramaniah L M and Nair S V 1993 *Phys. Rev.* B **47** 7132
4. Kuno M 2012 *Introductory Nanoscience* (Garland Science, London and New York)
5. Migdal A B 1959, in book of Landau L D and Lifshitz E M 2007 *Quantum Mechanics: Non-Relativistic Theory* (Elsevier, Singapore, Pte. Ltd.)
6. Warda K 2012 *Journal of Nanoscience and Nanotechnology* **12** 284
7. Shabaev A and Efros Al L 2004 *Nano Letters,* **4** 1821
8. Dvoyan K G, Hayrapetian D B and Kazaryan E M 2009 *Nanoscale Res. Lett.* **4** 106
9. Dvoyan K G, Kazaryan E M and Sarkisyan H A 2010, in Book *Modern Optics and Photonics*, Editors: Kryuchkyan G Yu, Gurzadyan G G, Papoyan A V (World Scientific)
10. Dvoyan K G, Hayapetyan D B, Kazaryan E M and Tshantshapanyan A A 2007 *Nanoscale Res. Lett.* **2** 601
11. Hayrapetian D B 2007 *Journal of Contemporary Physics* **42** 292
12. Hayrapetian D B, Dvoyan K G and Kazaryan E M 2007 *Journal of Contemporary Physics* **42** 151
13. Cantele G, Ninno D and Iadonisi G 2001 *Nano Letter* **1** 121
14. Cantele G, Ninno D and Iadonisi G 2001 *Phys. Rev.* B **64** 125325
15. Cantele G, Piacente G, Ninno D and Iadonisi G 2002 *Phys. Rev.* B **66** 113308
16. Leon H, Marin J L and Riera R *Physica* 2005 E **27** 385
17. Bagga A, Ghosh S and Chattopadhyay P K 2005 *Nanotechnology* **16** 2726
18. Gusev A A, Chuluunbaatar O, Vinitski S I, Dvoyan K G, Kazaryan E M, Sarkisyan H A, Derbov V L, Klombotskaya A S and Serov V V 2012 *Yadernaya Fizika* **75** 1281
19. Pöschl G and Teller E 1933 *Zs. Phys.* **83** 143
20. Komarov I V, Ponomarev L I and Slavianov S Yu 1976 *Spheroidal and Coulomb Spheroidal function* (Moscow: Nauka) (in Russian)
21. Kereselidze T M, Machavariani Z S and Noselidze I L 1998 *J. Phys. B: At. Mol. Opt. Phys.* **31** 15
22. Janke E, Emde F and Lösch F 1960 *Tafeln Höherer Funktionen* (Taubner B G Verlagsgesell-schaft, Stuttgart)
23. Bateman H and Erdelyi A 1953 *Higher Transcendental Functions* (Mc Graw-Hill Book Company ING)
24. Kereselidze T M, Machavariani Z S and Noselidze I L 1996 *J. Phys. B: At. Mol. Opt. Phys.* **29** 257
25. Kereselidze T M 1984 *Sov. Phys. JETP,* **60**, 423
26. Kereselidze T M 1987 *J. Phys. B: At. Mol. Phys.* **20**, 1891
27. Devdariani A, Kereselidze T M, Noselidze I L, Dalimier E, Sauvan P, Angelo P and Schott R 2005 *Phys. Rev.* A **71** 022512

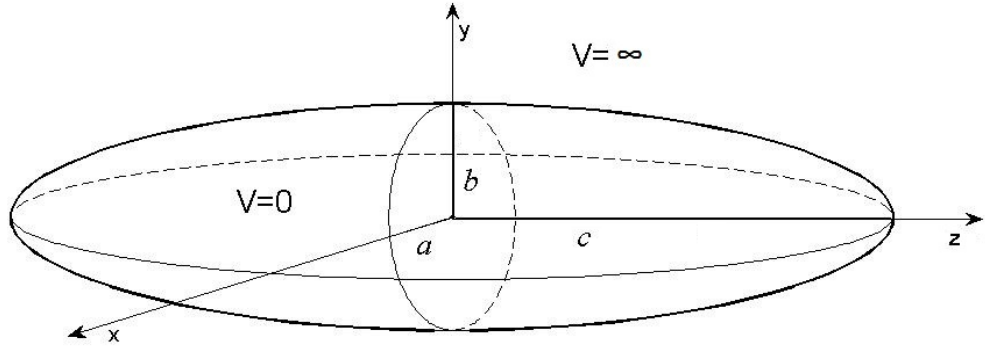

**Fig.1** Nanorod of ellipsoidal shape with major semi-axis $c$ (the nanorod axis) much larger than minor semi-axes $a = b$.

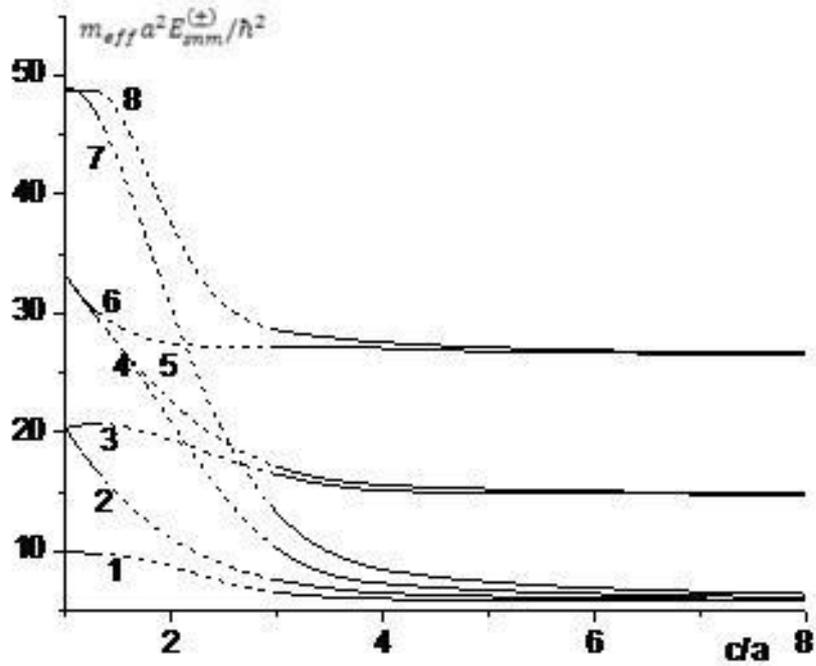

**Fig.2** The energy terms of an electron confined in nanorod of prolate ellipsoidal shape as a function of $c/a$. The energy terms are calculated with fixed $a$ and are shown in units of $\hbar^2 / 2m^* a^2$. The results are obtained using expression (20) for $c > 3a$ and expression (30) for $c \simeq a$. Legend: curve 1 – (100), $[100]^{(+)}$; curve 2 – (110), $[110]^{(-)}$; curve 3 – (111), $[101]^{(+)}$; curve 4 – (120), $[110]^{(+)}$; curve 5 – (121), $[111]^{(-)}$; curve 6 – (122), $[102]^{(+)}$; curve 7 – (130), $[120]^{(-)}$; curve 8 – (132), $[112]^{(-)}$; Sign $(\pm)$ in $[s,n,m]^{(\pm)}$ defines symmetry of the state.